\documentstyle[12pt]{article}
\begin{document}

\begin{titlepage}
\begin{center}

\bigskip

\bigskip

\vspace{3\baselineskip}

{\Large \bf  Thick Brane Worlds Arising From Pure Geometry}

\bigskip

\bigskip

\bigskip

\bigskip

{\bf Olga Arias, Rolando Cardenas and Israel Quiros}\\
\smallskip

{ \small \it  
Physics Department. Las Villas Central University.\\
Santa Clara 54830. Villa Clara. Cuba}

\bigskip

{\tt rcardenas@mfc.uclv.edu.cu}, {\tt  israel@mfc.uclv.edu.cu} 

\bigskip

\bigskip

\bigskip

\bigskip

\vspace*{.5cm}

{\bf Abstract}\\
\end{center}
\noindent
We study a non-Riemannian modification of 5-dimensional Kaluza-Klein theory. In our proposal the Riemannian structure of the five-dimensional manifold is replaced by a Weyl-integrable one. In this context a 4-dimensional Poincar$\grave{e}$ invariant solution is studied. Spacetime structures with thick smooth branes separated in the extra dimension arise. We focus our attention, mainly, in a case where the massless graviton is located in one of the thick branes at the origin, meanwhile the matter degrees of freedom are confined to the other brane. Due to the small overlap of the graviton's wave-function with the second thick brane, the model accounts for a resolution of the mass hierarchy problem a la Randall-Sundrum. Although, initially, no assumptions are made about the topology of the extra dimension, the solution found yields an extra space that is effectivelly compact and respects $Z_2$ symmetry.  Unlike some other models with branes, a preliminary study (not including the effect of scalar perturbations) shows that the spectrum of massive Kaluza-Klein states is quantized and free of tachyonic modes.


\bigskip

\bigskip

\end{titlepage}

\section{Introduction}

In the last 3 years increasing interest has been manifest on higher-dimensional space-times with large extra dimensions\cite{ah,rs,mass}. The ability of the proposed scenarios to answer the question: why the extra dimension is not observable?, rests on the brane construct; the standard model (SM) particles are confined to a three-brane embedded in the higher-dimensional "bulk" space-time, meanwhile gravity can propagate in the bulk. With the help of the brane scenarios some disturbing problems of high-energy physics, notably the mass hierarchy problem, can be succesfully treated\cite{mass}. Some difficulties with fine tunning and recovering of standard Friedmann cosmology on the brane seem to be the most undesirable features of these scenarios, although some alternatives have been investigated\cite{fc}.

The role of scalar fields in brane models has been extensivelly discussed in the bibliography\cite{sbs,bf,reall}. Usually scalars play the role of the modulus for the extra dimension, i. e., they are bulk fields\cite{bf}. In recent papers, however, the role of scalars in modeling thick (smooth) brane structures has been pointed out\cite{dewolfe,gremm}. In this case, although we have gravity coupled to scalars, these fields do not play the role of bulk fields but they provide the "material" from which the thick branes are made of.

An interesting feature of thick brane models based on gravity coupled to scalars is that one can obtain these space-time structures (branes) naturaly, i. e., without introducing them by hand in the action of the theory (by means, for instance, of delta functions)\cite{dewolfe}. 

The aim of the present work is to generalise former models based on gravity coupled to scalars by considering a non-Riemannian modification of 5-dimensional KK-theory. In our proposal the Riemannian structure of the 5-dimensional manifold is replaced by a Weyl-integrable one. Weyl-integrable geometry is a generalisation of Riemann geometry to allow for possible variations in the lenght of vectors during parallel transport. Riemann geometry is recovered when the scalar(Weyl)field is "frozen". One of the most exciting features of Weylian structures is that these mimic classically quantum behavior\cite{london}. These are, precisely, the main reasons why we use these structures in the present paper. 

In the present set-up the 4-dimensional scalar matter degrees of freedom and the massless graviton arise from pure geometry in five dimensions due to breaking of invariance of the theory under Weyl rescalings. Due to the simplicity of the model, we don't expect to recover the rich structure of matter of the standard model, instead we expect to show how, in a simple setup, several space-time structures with thick branes occur, in which the graviton and scalar matter degrees of freedom can be accommodated. The model we want to present is not intended to be a fundamental theory. Our action is purely 5d and, we expect, it could be the result of compactification of a more fundamental theory to five dimensions. 

The picture that emerges from the description we are about to present is exciting. If one imposes 4-dimensional Poincar$\grave{e}$ invariance, the structure of the 5-dimensional Weyl-integrable manifold $M_5^W$, that is consistent with the equations of our formalism, is non-trivial. The four-dimensional graviton and the scalar matter degrees of freedom are confined to several smooth thick branes with some separation in the extra dimension, depending on the region in the parameter space of the theory. The solution found respects, also, orbifold symmetry ($y\rightarrow -y$). 

In this paper we draw our attention to a particular case in which two thick branes are obtained; one with negative tension and the other with a positive one. In this case the graviton is concentrated near the origin in the extra coordinate, meanwhile the scalar matter is located in the second brane some distance away. The small overlap of the two "diffuse" thick branes accounts for a resolution of the mass hierarchy problem a la Randall-Sundrum (the precise form in which this mechanism works is not explored in detail in this paper). One of the most remarkable results is that, although we make no initial assumptions on the topology of the extra dimension, all the fields in the theory (including the massive KK-excitations) are confined to live within a subspace shaped by a potential well with infinite (rigid) walls. Therefore, $M_5^W$ is effectively compact. Besides, the spectrum of massive KK-modes is discrete (quantized). There exists a gap between the massless KK-mode (properly the graviton) and the lowest KK-excitation. In consequence, 4-dimensional gravity can be recovered by properly choosing one of the two free parameters of the theory. A preliminary study, not including perturbations of the scalar field, shows that the solution found is free of tachyonic modes so it is stable against small perturbations of the background metric.

In what follows Riemannian magnitudes and operators, i. e., those which are defined in respect to the Christoffel symbols of the metric $\{_{AB}^{\;C}\}=\frac{1}{2}g^{CN}(g_{AN,B}+\cdots)$, are denoted with a hat. Then, for instance: $\hat R_{AB}=\{_{AB}^{\;N}\}_{,N}-\{_{NA}^{\;N}\}_{,B}+\cdots$, $\hat\nabla_A V_B=V_{B,A}-\{_{AB}^{\;N}\}V_N$, etc. Upper case latin indexes $A,B,\cdots=0,1,2,3,5$, meanwhile, lower case ones $a,b,\cdots=0,1,2,3$. We use the following signature of the metric $g_{AB}=(-,+,+,+,+)$.

\section{The Model}

We will study a non-Riemannian modification (generalization) of Kaluza-Klein theory that is based on the following pure five-dimensional action,

\begin{equation}
\label{action}
S_5^W =\int_{M_5^W}\frac{d^5x\sqrt{|g|}}{16\pi G_5}e^{-\frac{3}{2}\omega}\{R+3\xi(\nabla\omega)^2+6U(\omega)\}, 
\end{equation}
where $M_5^W$ is a 5-dimensional Weyl-integrable manifold specified by the pair $(g_{AB},\omega)$, $g_{AB}$ being a 5-dimensional metric and $\omega$- a Weyl (scalar) function. The Weylian Ricci tensor $R_{AB}=\Gamma_{AB,N}^N-\Gamma_{NA,B}^N+\cdots$, where $\Gamma_{AB}^C=\{_{AB}^{\;C}\}-\frac{1}{2}(\omega_{,A}\delta_B^C+\omega_{,B}\delta_A^C-g_{AB}\omega^{,C})$ are the affine connections of $M_5^W$. $\xi$ is a coupling parameter and $U(\omega)$ is a "self-interaction" potential for $\omega$. In general, the self-interaction potential breaks invariance of the action (\ref{action}) under Weyl rescalings,

\begin{eqnarray}
\label{weylrescalings}
g_{AB}\rightarrow\Omega^2 g_{AB},\;\omega\rightarrow\omega+\ln\Omega^2,\nonumber\\
\xi\rightarrow\xi/(1+\partial_\omega\ln\Omega^2)^2,
\end{eqnarray}
where $\Omega^2$ is a smooth function on $M_5^W$. $U(\omega)=\lambda e^{-\omega}$, where $\lambda$ is a constant parameter, is the only functional form in which this potential does not break scale invariance of (\ref{action}). Breaking of Weyl invariance is the mechanism through which the Weyl scalar is transformed into physically observable matter degrees of freedom from which the smooth branes we obtain are made out. I. e., $\omega$ is not a bulk field playing the role of the modulus for the extra dimension.

We will look for solutions to the theory (\ref{action}) that respect 4-dimensional Poincar$\grave{e}$ invariance, and we propose the following (warped) ansatz for the line-element:

\begin{equation}
\label{linee}
ds_5^2=e^{2A(y)}\eta_{nm}dx^n dx^m+dy^2,
\end{equation}
where $e^{2A}$ is the "warp" factor, $\eta_{ab}$ is the (flat) Minkowski metric, and $y$ stands for the extra coordinate. We will make no assumptions on the topology of the extra dimension so, in principle, $-\infty<y<+\infty$. In this sense our set-up is much like to the one by Wesson\cite{wesson}. Recall that (\ref{action}) is of pure geometrical nature since, the scalar function $\omega$ enters in the definition of the manifold affine connections.

We expect that, in this particular case, thick brane structures should arise since similar 5-d actions with scalar fields (but on Riemannian manifolds) yielding thick brane 5-d spacetimes have been studied, for instance, in Ref.\cite{dewolfe,gremm}. The nobel feature here is that, spacetime structures with several (one, two, four, etc.) thick smooth branes arise, depending on the region one chooses in parameter space of the solution.

\section{The Solution}

In the search for a 4-dimensional Poincare-invariant solution to the theory given by equations (\ref{action}) and (\ref{linee}), we will use the conformal technique, which beneficts have been pointed out, for instance, in Ref. \cite{faraoni}. Under the conformal transformation $\hat g_{AB}= e^{-\omega} g_{AB}$, the Weylian affine connections $\Gamma_{AB}^{\;C}\rightarrow\hat{\{_{AB}^{\;C}\}}$ and, in general, $M_5^W\rightarrow M_5^R$- a five-dimensional manifold of Riemannian structure. Besides,

\begin{equation}
\label{confaction}
S_5^W\rightarrow S_5^R=\int_{M_5^R}\frac{d^5x\sqrt{|\hat g|}}{16\pi G_5}\{\hat R+3\xi(\hat\nabla\omega)^2+6\hat U(\omega)\},
\end{equation}
where $\hat U(\omega)=e^\omega U(\omega)$ and, as pointed out in Sect. {\bf 1}, hatted magnitudes and operators refer to Riemann structures. The action (\ref{confaction}) has been studied recently in thick brane contexts\cite{dewolfe,gremm} and, in general, a similar action but with domain walls explicitely introduced, has been investigated fairly extensively in Ref.\cite{reall}. In this case, however, the scalar field plays the role of a modulus field for the extra dimension (it is a bulk field)\cite{reall}. We recall that, in the case presented here, the scalar field is not a modulus field. What we want is, precisely, to obtain smooth brane structures without introducing them explicitely in the action.

Under $g_{AB}\rightarrow e^\omega\hat g_{AB}$, the line-element (\ref{linee}) is mapped into,

\begin{equation}
\label{conflinee}
ds_5^2=e^{2\sigma(y)}\eta_{nm}dx^n dx^m+e^{\omega(y)}dy^2,
\end{equation}
where, $2\sigma=2A+\omega$. If one introduces new variables $X\equiv\omega'$ and $Y\equiv 2A'$ (the comma denotes y-derivative), then the field equations that are derivable from (\ref{confaction}), taking into account the ansatz (\ref{conflinee}), reduce to the following pair of coupled equations:

\begin{eqnarray}
\label{fielde}
X'+2YX-\frac{3}{2}X^2=\frac{1}{\xi}\frac{d\hat U}{d\omega}e^{-\omega},\nonumber\\
Y'+2Y^2-\frac{3}{2}XY=(\frac{1}{\xi}\frac{d\hat U}{d\omega}+4\hat U)e^{-\omega}.
\end{eqnarray}

This set of equations can be easily solved if one uses the symmetry of the left-hand side (LHS) of the above equations under the restriction $X=k\;Y$, where $k$ is some constant parameter. This restriction is, in fact, a constrain on the kind of potentials for which our solutions hold: 

\begin{equation}
\label{selfintpot}
\hat U=\lambda e^{\frac{4k\xi}{1-k}\omega}.
\end{equation}

After the restriction $X=k\;Y$ and taking into account (\ref{selfintpot}), both equations in (\ref{fielde}) reduce to a single differential equation,

\begin{equation}
\label{finale}
Y'+\frac{4-3k}{2}Y^2=\frac{4\lambda}{1-k}e^{(\frac{4k\xi}{1-k}-1)\omega}.
\end{equation}

There is a particular case in which this equation is easily solvable. If we choose $\xi=\frac{1-k}{4k}$, then the self-interaction potential in the action (\ref{action}) $U=\lambda$, obviusly breaks invariance under Weyl rescalings (including conformal symmetry). This choice is further responsible for transformation of the geometrical Weyl scalar $\omega$ into an observable scalar field. In what follows we will study this symmetry breaking potential. As a consequence the field equation (\ref{finale}) simplifies to,

\begin{equation}
\label{finale1}
Y'+\frac{4-3k}{2}Y^2=\frac{4\lambda}{1-k}.
\end{equation}

Subsequent integration of (\ref{finale1}) yields,

\begin{equation}
\label{firstintegral}
Y=ab \tanh(ay),
\end{equation}
where,

\begin{equation}
\label{constants}
a=\sqrt{\frac{4-3k}{1-k}2\lambda},\;\;b=\frac{2}{4-3k}.
\end{equation}

In consequence, the solution to our set-up is given by the following expresssions:

\begin{equation}
\label{warpf}
e^{2A(y)}=[\cosh(ay)]^b,
\end{equation}

\begin{equation}
\label{scalar}
\omega=kb\ln\cosh(ay).
\end{equation}

As seen from (\ref{warpf}) and (\ref{scalar}), the solution found respects, besides, $Z_2$ symmetry ($y\rightarrow -y$). This means that, in the $y$-direction, one half of the manifold (say the lower half) is a mirror reflection of the other one and, consequently, only this part of the manifold is physically relevant.

We should stress that we have, in fact, a class of solutions depending on the region in parameter space:

(A) $\lambda>0$, $k>4/3$

In this case, $a\in \Re$ and $-\infty<y<+\infty$, i. e., we have a non-compact manifold in the extra-dimension (we recall that, due to orbifold symmetry of the solution, only one half of the extra dimension, say $0\leq y<+\infty$, is physically relevant). On the other hand $b<0$ is negative and the warp factor (in fact the wave-function of the graviton up to a positive power of $b$ as it will be shown later on), is concentrated near of the origin $y=0$ as seen in Fig. 1. Since the 5d stress-energy tensor is given by,

\begin{equation}
\label{stresse}
T_{AB}=\frac{1}{8\pi G_5}(R_{AB}-\frac{1}{2}g_{AB}R),
\end{equation}
then, its 4d and pure 5d parts are given through the following expressions,

\begin{equation}
\label{4dstresse}
T_{ab}=\frac{3\lambda}{8\pi G_5}[\cosh(ay)]^b\{1+\frac{k}{4-3k}\tanh^2(ay)\}\eta_{ab},
\end{equation}
and,

\begin{equation}
\label{5dstresse}
T_{55}=\frac{3\lambda}{8\pi G_5}\{1-\frac{k}{4-3k}\tanh^2(ay)\},
\end{equation}
respectively. The energy density of the scalar matter (in fact the null-null component of $T_{ab}$ in Eq. (\ref{4dstresse})) looks like,

\begin{equation}
\label{mu}
\mu(y)=-3\lambda[\cosh(ay)]^b\{1+\frac{k}{4-3k}\tanh^2(ay)\}.
\end{equation}

For $4/3<k<2$ this energy-density function shows a minimum at $y=0$ where $\mu$ is negative and a maximum at some $y\neq 0$ where $\mu$ is positive. It asymptotes zero for $y\rightarrow\pm\infty$ (see Fig. 2). The fact that we have both negative and positive values for $\mu$ can be compared with the standard (Randall-Sundrum) thin brane case, where one of the branes has a positive brane tension meanwhile the second brane has a negative one\cite{rs}. The 5d curvature scalar,

\begin{equation}
\label{5dricci}
R_5=-2\lambda\{5+\frac{3k}{4-3k}\tanh^2(ay)\},
\end{equation}
is always bounded as seen from Fig. 3, meaning that the five-dimensional manifold is non-singular.

(B) $\lambda>0$, $1<k<4/3$

In this case, $a\in \Im$, $b>0$. Consequently, in the above equations we should replace $a\rightarrow i\bar a$, $\cosh(ay)\rightarrow\cos(\bar a y)$ and $\tanh(ay)\rightarrow i\tan(\bar a y)$, where $\bar a=\sqrt{\frac{4-3k}{k-1}2\lambda}$, i. e., we have a manifold that is periodic in the $y$-direction so $-\pi\leq\bar a y\leq\pi$ and we can treat the compact case. Besides, since our solution respects $Z_2$ symmetry, we can take the physically relevant part of the manifold to be $0\leq y\leq\pi/\bar a$. The shape of the warp factor is shown in Fig. 4.

The 4d and pure 5d parts of the 5d stress-energy tensor (\ref{stresse}) are given through the following expressions,

\begin{equation}
\label{4dstresse'}
T_{ab}=\frac{3\lambda}{8\pi G_5}[\cos(\bar a y)]^b\{1-\frac{k}{4-3k}\tan^2(\bar a y)\}\eta_{ab},
\end{equation}
and,

\begin{equation}
\label{5dstresse'}
T_{55}=\frac{3\lambda}{8\pi G_5}\{1+\frac{k}{4-3k}\tan^2(\bar a y)\},
\end{equation}
respectively. The energy density of the scalar matter is therefore,

\begin{equation}
\label{mu'}
\mu(y)=-3\lambda[\cos(\bar a y)]^b\{1-\frac{k}{4-3k}\tan^2(\bar a y)\}.
\end{equation}

The shape of the scalar energy density is shown in Fig. 5. It shows a series of smooth thick branes with both positive and negative energy densities interpolated between the boundaries of the manifold. The ones at $y=0$ and $y=\pi/\bar a$ are of negative energy density, meanwhile, there are two smooth branes with positive energy density located in between them. The 5d curvature scalar,

\begin{equation}
\label{5dricci'}
R_5=-2\lambda\{5-\frac{3k}{4-3k}\tan^2(\bar a y)\},
\end{equation}
is singular at $y=\pi/2\bar a$ (see Fig. 6).

The other physically relevant case: $\lambda<0$, $k>4/3$ coincides with case (B). Other possible situations, for instance: $\lambda>0$, $k<1$, are not physically relevant since, in this case, $a\in\Re$ and $b>0$, meaning that the graviton wave function is a minimum at $y=0$ and infinitely increases while $y\rightarrow\infty$. In other words, it is not normalizable.

\section{Kaluza-Klein Modes}

We now turn to the question about the chance for our set-up to describe 4-dimensional gravity. For this purpòse we will examine fluctuations of the metric around the classical background solution (\ref{linee}), (\ref{warpf}),

\begin{equation}
\label{mfluct}
ds_5^2=e^{2 A(y)}(\eta_{nm}+\epsilon h_{nm}(x,y))dx^n dx^m+dy^2,
\end{equation} 
where $\epsilon<<1$. In general, in gravity coupled to scalars, one can not avoid fluctuations of the scalar while treating fluctuations of the background metric. However, mathematical treatment of these coupled fluctuations is rather complicated. In this case it is used to treat a sector of the metric fluctuations that decouples from the scalar and metric fluctuations can be studied analytically \cite{gremm}. This simplification is based on a method developed in Ref. \cite{dewolfe}. Following this method, we consider only the transverse traceless components of $h_{ab}$. Besides, as before, we "jump" to the conformal frame ($g_{AB}\rightarrow e^\omega\hat g_{AB}$), where the field equations are simpler. We are left with the following wave-equation for the metric fluctuations,

\begin{equation}
\label{wavee}
(\partial_z^2+4\sigma'\partial_z+e^{-2\sigma}\Box^\eta)\hat h_{ab}^T=0,
\end{equation}
where $\hat h_{ab}^T$ is the transverse, traceless mode that decouples from the scalar. Besides, we have introduced a new $z$-coordinate: $dz=e^{-\frac{\omega}{2}}dy$, and $\Box^\eta$ is the (flat) Minkowski wave operator. In Ref. \cite{dewolfe}, it has been shown that the above equation supports a naturally massless and normalizable 4-dimensional graviton given by $\hat h_{ab}^T=C_{ab}e^{ipx}$, where the $C_{ab}$ are constants and $p^2=0$. Going back to the original frame ($h_{ab}^T=e^\omega \hat h_{ab}^T$), we obtain for the massless graviton of our set-up: $h_{ab}^T(x,y)=C_{ab}\psi_0(y)e^{ipx}$, where $\psi_0(y)=e^{2kA}=[\cosh(a y)]^{kb}$, is the wave-function of the graviton in the extra dimension. The fact that this wave-function, in the case (A) of Sec. 3 ($\lambda>0$, $k>4/3$), is normalizable will be evident later on, when we show that the five-dimensional manifold we are dealing with is effectively compact in this case. For the sake of brevity, the other (physically less interesting) cases of our solution, including case (B) of Sec. 3, will not be studied in this section. For the massive KK-excitations we obtain the following wave equation in Schr$\ddot{o}$dinger form:

\begin{equation}
\label{shroedinger}
(\partial_r^2+[m^2-V(r)])\psi=0,
\end{equation}
where we have dropped the subscripts in $\psi$, $m$ is the mass of the KK-excitation and the following coordinate change has been considered: $dr=e^{-\sigma}dz$. The shape of the potential $V(r)$ in the case of interest here ($\lambda>0$, $k>4/3$) is given by the following expression:

\begin{equation}
\label{potential}
V(r(y))=3\lambda[\cosh(a y)]^{-b}[1+\frac{3k-5}{2(4-3k)}\tanh^2(a y)].
\end{equation}

Much can be extracted from the analysis of the form of this potential, even without trying to solve explicitely Eq. (\ref{shroedinger})\footnotemark\footnotetext{It is behind the scope of the present work to solve explicitely Eq. (\ref{shroedinger}). A first intent to solve it analitically was unsuccesful.}. In effect, Eq. (\ref{shroedinger}) with the potential given by Eq. (\ref{potential}), is an analog of a quantum mechanics situation with a potential well of infinite (rigid) walls (see Fig. 7). Therefore, unlike the original thin-brane models \cite{rs,mass,fc} and some other thick-brane models \cite{dewolfe,gremm}, the spectrum of massive KK-excitations is discrete. Consequently, there is a gap between the massless (graviton) mode and the first excited KK-mode. Hence, by properly choosing the free parameter $\lambda$, we can put the first massive KK-excitation out of the reach of the low-energy experiments and so, we are able to recover 4-dimensional gravity in our set-up. A rather crude estimate of the situation discussed here could be given if we approach the complicate potential $V(r)$ in Eq. (\ref{potential}) by a parabolic potential $V(r)\sim y^2$, typical in the quantum treatment of the linear harmonic oscillator. In this case the mass spectrum is given by $m_n=\bar m_0\sqrt{n+\frac{1}{2}}$, where $n=0,1,2,\cdots$. The mass of the first massive KK-excitation $m_0=\bar m_0/\sqrt{2}$ is a function of $\lambda$. With increasing $\lambda$, $m_0$ increases (the shape of the potential well is narrower as shown in Fig. 7). Since the mass $m\sim\sqrt{n+\frac{1}{2}}$ then, a continuum KK-spectrum is obtained in the $n$-large limit (the classical limit). Although, only by solving exactly the Schr$\ddot{o}$dinger equation (\ref{shroedinger}) one could find the exact dependence of $m$ on the parameter $\lambda$, i. e., to obtain by inversion $\lambda=\lambda(n)$, one can speculate that in the $n$-large limit a thin-brane picture could be recovered.

Finally, we want to stress that all the KK-modes in our set-up, are non-spacelike so, our solution is tachyon-free, thus warranting its stability. Besides, due to the shape of the potential well $V(r)$ in Eq. (\ref{potential}) (see Fig. 7), all the matter degrees of freedom, including the KK-modes, are confined to a subspace in $M_5^W$ whose boundaries are fixed by $V(r(y))$\footnotemark\footnotetext{The wave function for the KK modes are different from zero only within the potential well}. In consequence, all the KK-states in our set-up (including the massles "graviton" mode) are normalizable. We recall, however, that our study is just a preliminary one since a more accurate treatment should include fluctuations of the scalar field $\omega$ as well.

\section{Conclusions and Comments} 

In this conclusive Section we will just enumerate the relevant features of the present model and, briefly, to comment on them.

\begin{itemize}

\item Since we start from the action (\ref{action}), that has pure geometrical nature (including the Weyl symmetry breaking term $U=\lambda$), the graviton and scalar matter degrees of freedom, being accomodated within a five-dimensional manifold with the non-trivial structure discussed above, arise from pure geometry in $M_5^W$.

\item In the subspace of our solution $\lambda>0$, $k>4/3$ (case (A) of Sec. 3), the non-trivial structure of $M_5^W$ emerging from our set-up, in which the graviton wave-function $\psi_0(y)=e^{2kA}=[\cosh(a y)]^{kb}$ is concentrated near of the origin, and has a small overlap with the mass density function $\mu(y)$ (which is a maximum at some $y\neq 0$), can take account for a resolution of the mass hierarchy problem a la Randall-Sundrum\cite{mass}. However, this point deserves further investigation since, it is not clear which criterium to use for exactly fixing the overlapping point in $y$-coordinate, where the correct hierarchy is generated. One criterium could be to take as the intersection point the value $y$ where $\mu(y)$ is a maximum. However, other criteria may be discussed.

\item Since the KK-spectrum supported by the potential $V(r(y))$ in Eq. (\ref{potential}) is quantized, i. e., there exists a gap between the massless (graviton) mode and the first excited KK-mode, we can succesfully describe 4-dimensional (low-energy) gravity by a proper choice of the free parameter $\lambda$. However, in the present paper we have not atempted at such a proper choice. It could be desirable that this choice would be consistent with generation of the correct mass hierarchy (see the former point).

\item It is nice that there are not tachyon modes in the KK-spectrum so, the solution found is stable against small perturbations of the background metric. However, this point should be further investigated since we have not considered perturbations of the scalar (Weyl) function $\omega$ as it could be.

\item The fact that a discrete (quantized) spectrum of KK-states arises naturally (espontaneously), could be taken as just another confirmation of the claim that Weylian structures mimic, classicaly, quantum behavior\cite{london}. However, it could be of interest to investigate whether the quantized KK-spectrum, naturally obtained in the present model, is really a consequence of the Weylian structure assumed for our 5-dimensional manifold $M_5^W$, or is due to another considerations in the present set-up. We could, for instance, to start with the present action (\ref{action}) but considering Riemannian manifolds $M_5^R$. In this case the action (\ref{action}) is just the low-energy (dilatonic) string action in five dimensions\cite{wands}.

\item One of the most remarkable features of our set-up is that it contains (naturally) a mechanism of effective (spontaneous) compactification of the extra space that could be further investigated.

\end{itemize}

Finally, we want to enumerate some points that deserve further investigation. First, other solutions to our model can be found. For instance, by using the generosity of conformal techniques, one can find solutions that are conformal to the ones encountered in the bibliography to the action (\ref{confaction})\cite{dewolfe,gremm}. These represent solutions to our set-up as well, but with different self-interaction potentials (possibly breaking the original symmetry of the theory under the Weyl rescalings too). Second, it could be desirable to solve explicitely the Schr$\ddot{o}$dinger equation (\ref{shroedinger}) with the potential (\ref{potential}). This could yield more insight into the present model and its possible modifications. Finally, it could be of interest to extend our set-up to higher dimensions.


We acknowledge the MES of Cuba by financial support of this research.


\newpage

\begin{figure}[b]
\caption{The shape of the warp factor $e^{2 A(y)}$ is shown for arbitrary $\lambda=0.00001$ and $k > 4/3$. Since this is in fact the wave function of the graviton in the extra dimension, it is seen that the graviton is concentrated near of the origin $y=0$} 
\end{figure}

\begin{figure}[b]
\caption{Shape of the scalar energy density function $\mu$ for arbitrary $\lambda=0.00001$ and $4/3 < k < 2$. A thick brane with negative energy density is located near of the origin $y=0$ (the graviton one) and by this brane the one with positive energy density is seen (the one where the scalar matter is concentrated).} 
\end{figure}

\begin{figure}[b]
\caption{The functional form of the curvature scalar $R_5$ for arbitrary $\lambda=0.00001$ and $k > 4/3$ is shown. It is always bounded and well behaved.} 
\end{figure}

\begin{figure}[b]
\caption{The shape of the warp factor $e^{2 A(y)}$ is shown for arbitrary $\lambda=0.00001$ and $1 < k < 4/3$. Periodicity in $y$ and orbifold symmetry are appreciated} 
\end{figure}

\begin{figure}[b]
\caption{Shape of the scalar energy density function $\mu$ for arbitrary $\lambda=0.00001$ and $1 < k < 4/3$. Two thick branes with negative energy density are located at the boundaries of the manifold $y=0$ and $y=\pi/\bar a$, meanwhile two positive energy branes are located in between them.} 
\end{figure}

\begin{figure}[b]
\caption{The functional form of the curvature scalar $R_5$ for arbitrary $\lambda=0.00001$ and $1 < k < 4/3$ is shown. A singularity is appreciated at $y=\pi/2\bar a$.} 
\end{figure}

\begin{figure}[b]
\caption{The shape of the potential $V(r(y))$ is shown for two distinct values of the parameter $\lambda$: $\lambda_1=0.00001$ (dotted line) and $\lambda_2=0.0001$ (solid line). It is seen that with increasing $\lambda$ the shape of the potential well becomes narrower} 
\end{figure}

\end{document}